# Applications of artificial intelligence in drug development using real-world data

Zhaoyi Chen[1*], Xiong Liu[2*], William Hogan[1], Elizabeth Shenkman[1], Jiang Bian[1]

1. Department of Health Outcomes and Biomedical Informatics, College of Medicine, University of Florida
2. AI Innovation Center, Novartis

* These two authors contributed equally to this work and are co-first authors.

Corresponding author: Jiang Bian, PhD (e-mail: bianjiang@ufl.edu)

**Teaser**: This paper highlights the progress and challenges in current drug development studies using both artificial intelligence and real-world data.

**Highlights**:
- Artificial intelligence (AI) and real-world data (RWD) are increasingly used in drug development
- Adverse event detection, trial recruitment, and drug repurposing are the most popular tasks
- Significant gaps in current studies include data quality, casual inference, and transportability
- Major national and international initiatives are making RWD increasingly available
- Novel distributed analytical strategies are needed to facilitate the use of RWD

**Abstract**
The US Food and Drug Administration (FDA) has been actively promoting the use of real-world data (RWD) in drug development. RWD can generate important real-world evidences reflecting the real-world clinical environment where the treatments are used. Meanwhile, artificial intelligence (AI), especially machine-/deep-learning (ML/DL) methods, have been increasingly used across many stages of the drug development process. Advancements in AI have also provided new strategies to analyze large, multi-dimensional RWD. Thus, we conducted a rapid review of articles from the last 20 years, aiming to provide an overview of the drug development studies that use both AI and RWD. We found the most popular applications are adverse event detection, trial recruitment, and drug repurposing. We discussed current research gaps and future opportunities.

**Introduction**
Drug development is the process of bringing a new drug molecule into clinical practice; in its broadest definition, it includes all stages from the basic research of finding a suitable molecular target to large-scale Phase III clinical studies that support the commercial launch of the drug to post-market pharmacosurveillance and drug repurposing studies [1,2]. During the drug development process, chemical entities that have the potential to become therapeutic agents will have to be identified and thoroughly tested, and the entire process is lengthy and costly. It is estimated that for every new drug brought to the market, it typically costs billions of U.S. Dollars and more than 10 years of work [3,4]. Therefore, strategies that can facilitate and accelerate the drug development process are of high interest.

Recently, the FDA has been actively promoting the use of real-world data (RWD) for drug development [5,6]. The term 'RWD' refers to data collected from sources outside of conventional research settings, including electronic health records (EHRs), administrative claims, and billing data among others [5–7]. These RWD often contain detailed patient information such as disease status, treatment, treatment adherence and outcomes, comorbidities, and concurrent treatments that are tracked longitudinally. The information generated from RWD can provide important real-world evidence to inform therapeutic development, outcomes research, patient care, safety surveillance, and comparative effectiveness studies [8]. More importantly, the use of RWD allows clinical researchers and regulatory agencies to answer questions more efficiently, saving time and money while yielding answers that are generalizable to the broader population. Over the past decade, there has been an increased uptake of EHR systems in the USA. These technological advances and policy changes in the USA have created a fertile ground with increasing opportunities to use RWD to facilitate drug development. Thus, the FDA has provided guidance on the use of EHR data in clinical investigations [5] as well as guidance on incorporating RWD into regulatory submissions to the FDA [9].

By contrast, the field of AI, including machine learning and deep learning (ML/DL), has moved from largely theoretical studies to real-world applications thanks to both the exponential growth of computing power and advances in AI methods [10]. AI has been widely used in many stages of the drug development process to identify novel targets [11], increase understanding of disease mechanisms [12], and develop new biomarkers [13] among others. Many pharmaceutical companies have begun to invest in resources, technologies, and services, especially in generating and assembling datasets to support research in AI and ML/DL, and many of those datasets are from RWD sources. There is an emerging need for an overview of the intersection between AI and RWD in current drug development studies to describe the current trends, identify existing research gaps, and provide insights into potential future directions. Thus, we conducted a rapid review summarizing published articles related to the intersection of AI, RWD, and drug development over the past 20 years. Our specific aims were to identify: (1) what are the current trends of using AI and RWD in drug development studies? and, subsequently (2) what are the challenges and opportunities?

**Literature search**
*Definitions of drug development, AI, and RWD*
The drug development process, according to the FDA's definition [14], has four stages: (1) drug discovery: the discovery of new therapeutic agents through the understanding of disease mechanisms and properties of molecular compounds (or other technologies); (2) preclinical research: laboratory and animal testing to answer questions about the safety of the new drug targets; (3) clinical research: different stages of clinical trials to test the new drug on humans to assess its safety and efficacy; and (4) post-marketing research: pharmacosurveillance and comparative effectiveness studies.

The definition of AI methods is less clear and varies in computer science and informatics literature. In this rapid review, we chose the definition as "*the use of complex algorithms and software to emulate human cognition in the analysis of complicated medical data, and analyze the relationships between prevention or treatment techniques and patient outcomes.*" [15] To be more concrete, the specific AI-related methods we considered include machine learning (ML) and deep learning (DL), (a sub-branch of ML), which are in general accepted by different research communities as AI tasks [16].

In terms of RWD, the FDA defines RWD as "*the data relating to patient health status and/or the delivery of health care routinely collected from a variety of sources*," which include patients' EHRs and claims data, as well as other patient-generated health data such as those generated in home-use care settings and data from mobile devices that can inform health status [7,8]. In this paper, as we aim to understand RWD that can be used to support drug development, we focus on RWD sources that provide clinical data not collected in interventional, controlled, experimental clinical research settings (e.g., randomized controlled trials [RCTs]), which include data generated not only from the delivery of routine care (e.g., EHR, claims databases, or disease registries) but also from study designs that can generate RWD (e.g., observational studies and pragmatic clinical trials) [17]. We exclude RWD that are generated from personal devices such as smartphones and activity trackers.

*Eligibility criteria*
The inclusion criteria for our review were: (1) studies using RWD as data sources; (2) studies using AI methods for statistical analysis or data mining; and (3) studies focused on the development of drugs. As a rapid review, we first focused on identifying existing review articles.

*Search strategy and study selection*
We performed a literature search through PubMed to identify relevant review articles published until July 1, 2020. In our search strategy, we considered different combinations of search keywords dictated by the definitions of RWD, AI, and drug development that we chose to focus on. Our search query included three distinct sets of keywords for RWD, AI, and different stages of the drug development process, respectively. For completeness, we included keywords such as "*natural language processing*" in the AI keywords, because state-of-the-art models for these NLP tasks are often ML/DL methods. The full search query and the complete list of keywords are in the **Supplementary Table S1**.

Following best practice for rapid reviews [18,19], we first restricted our search to identify existing review articles for inclusion. We then manually identified the specific AI and RWD applications described in these reviews. Next, based on the identified applications, we performed a second round of literature search to look for their detailed approaches, including data source, data type, and analytical methods used. In *Figure 1*, we summarize the overall search and screening process.

**Current progress in the literature**
In the first round of literature search, a total of 23 review articles were identified; among them, 16 met our inclusion criteria. Based on these review papers, in *Figure 2 Panel 1*, we first highlight the key steps in the drug development process and then summarize the identified research topics in each step. We then summarize the applications that used RWD (*Figure 2 Panel 2*) and AI+RWD (*Figure 2 Panel 3*) to address these research questions.

*Drug development process and applications of real-world data (RWD)*
The first step in the drug development process is the discovery of potential therapeutic agents, where researchers investigate the interactions among different molecules, genes, and proteins, and then identify which molecules have high potential with the goal of finding novel targets, biomarkers, and compounds [14]. Some of these goals can be achieved using RWD applications. For example, in a recent review paper [17], Singh et al. identified 20 studies that used RWD to facilitate drug discovery and clinical research. Among them, 16 identified or validated new phenotypes, disease markers, and biomarkers for patient identification and stratification.

The next step is preclinical testing, which includes both *in vitro* and *in vivo* testing. In this stage, the safety of drug molecules is tested in test tubes, living cell cultures, and animal models. This is a crucial step because the drug development can only move into human trials with extensive data on safety in preclinical research. In the review papers we included, there were no studies identified for this stage.

After the preclinical testing, once the Investigational New Drug (IND) application is approved, drug development moves into clinical research stages. There are three phases of clinical studies before the drug can be submitted for marketing approval. The key issue that needs to be addressed in this step is to evaluate both the safety and efficacy of the new agents in the target human population [20]. Randomized controlled trials (RCTs) are still the gold standard to generate clinical evidence; however, RWD have become an important data source for RCTs to understand how the developed treatments are being used in real-world settings. For example, Lai et al. examined the impact of using EHRs for clinical research recruitment in a review of 13 research articles [21]. They found that the automation in screening and patient identification could contribute to higher recruitment yield and reduced workload.

After a drug is available on the market, the drug developers are required to submit regular reports detailing adverse events (AEs) associated with the drug [14]. In addition to AE reporting, observational studies and pragmatic clinical trials are also conducted using RWD to evaluate the safety of the drug in real world settings. For pharmacosurveillance, RWD has gained significant attention in recent years. For example, in 2012, Warrer et al. conducted a review on studies that used text-mining techniques on narrative documents to investigate adverse drug reactions [22], where only 7 studies were identified. In a more recent review conducted by Luo et al. in 2017 on the same topic, a total of 48 studies were identified [23]. These studies showed that text-mining techniques, ranging from simple free-text searching to more advanced ML/DL-based natural language processing (NLP) methods, can be powerful in adverse events detection, given that adverse events are more extensively documented in EHR narratives.

*Applications of AI methods using RWD in the drug development process*
Across the different drug development stages, few studies used AI on RWD, and most were found in the clinical or post-marketing stage. Three main types of study used AI on RWD (**Figure 2 Panel 3**): trial recruitment optimization, adverse events detection, and drug repurposing. Therefore, we conducted a second round of literature search focusing on individual research studies of these three main applications as shown in *Figure 1*. Similar to the first round of literature search, we screened all studies on these three topics using keywords related to AI and RWD as detailed in the Methods section. A total of 65 research studies were included after title/abstract and full-text screening. In **Table 1**, we summarize these studies into subcategories with examples. In *Figure 3*, we show the increasing trend of studies that use AI methods with RWD in the drug development process over the last 15 years. Overall, we observed a steady increase in the total number of studies. In particular, the number of studies focusing on adverse event detection has exploded and many focused on using NLP methods to extract adverse events from free-text narratives, likely because of advances in DL-based NLP methods that achieved state-of-the-art performance[24]. Nevertheless, we also observed more studies that tried to leverage AI methods on RWD for optimizing clinical trial recruitments. Moreover, clinical drug repurposing has emerged as a new application area in the drug development process.

**Figure 4** summarizes the numbers and percentages of different data sources, data types, and AI methods being used in the 65 studies. Given the overwhelming number of studies used AI-driven NLP methods, we separated NLP studies from other ML/DL studies. Note that the state-of-the-art NLP methods often leverages ML and DL approaches such as BERT[24,25]. Overall, EHR data were the most popular data source, especially unstructured clinical notes. Consequently, a large number of studies have focused on developing or using NLP methods. Among the 55 studies on AE detection, 41 (74.5%) of them were NLP-related. Some studies developed a NLP system to extract information from clinical notes to identify AEs related to the administration of medication. For example, Yang et al. developed a Long Short-Term Memory (LSTM)-based DL model to detect medication, AEs, and their relations from clinical text[26]. In other studies, the AEs and associated attributes (such as severity) extracted from the NLP pipeline were further fed into a downstream model to assess association between AEs and other health outcomes. For example, Zhang et al. first used NLP to identify patients who had AE related to statin therapy, and then examined the relationship between continuation of statin therapy and incidence of death and cardiovascular events among these patients [27]. Meanwhile, the majority of studies (75% of studies included) on recruitment optimization also utilized clinical notes from EHR data, and attempted to identify eligible populations for trials using information extracted from NLP. For example, Spasic et al. used an NLP system that combined rule-based knowledge infusion and machine learning algorithms to analyze longitudinal patient records to determine if the corresponding patients met given eligibility criteria for clinical trials [28]. Finally, for the two articles on clinical drug repurposing [29,30], one of them used NLP methods. In work by Xu et al., automated informatics methods including NLP were used on EHR data to identify patient cohorts and medication information[29], and they then assessed whether metformin is a potential drug that can be repurposed to cancer treatment. In the other clinical drug repurposing study, Kuang et al. developed a ML-based drug repurposing approach, called baseline regularization, to predict the effects of drugs on different physical measurements such as fasting blood glucose [30] to identify potential repurposing. Note that there is a wealth of literature on drug repurposing using EHRs; however, very few used advanced AI methods, while the majority uses traditional statistical approach such as Cox regression.[31]

*Current trends of AI methods on RWD in drug development research*

We identified 16 review articles related to the use of AI methods on RWD published over the past 20 years and an increasing number of original studies in three main application areas: adverse events detection, recruitment optimization, and drug repurposing.

The most common application area that used AI on RWD was for AE detection, primarily focusing on using NLP on unstructured clinical notes from EHR. The reasons for such a rising popularity is two-fold: (1) the abundance of textual information in RWD, especially EHRs, and (2) the rapid advancement in NLP methods, especially those new deep learning-based models with state-of-the-art performance. In fact, over 80% of the clinical information in EHR is documented in free-text [37], which makes text mining an ideal tool. EHRs have been particularly useful for investigating AEs and other therapeutic effects because of their continuous and longitudinal nature of clinically relevant outcomes and medication exposures.

We also identified several studies that focused on recruitment optimization and drug repurposing. These tasks are suitable for the use of AI and RWD because (1) the extensive collections of RWD provide sufficient sample sizes to identify individuals that meet recruitment criteria, (2) the longitudinal detailed medical histories of patients captured in these RWD sources make it possible for researchers to identify drugs that may be effective for indications other than the primary use, (3) AI and data-driven approaches could potentially minimize the selection bias as they do not rely on researchers' pre-determined assumptions, and thus, are able to identify novel associations that were previously unknown, and (4) modern AI methods are capable of handling the high dimensionality and complexity of RWD as well as the complex combinations and interactions of RWD variables.

**Challenges and future directions**
*Challenges of using AI and RWD in the drug development studies*

First, one major challenge is the quality of the data in many RWD sources. For example, information heterogeneity has been reported in EHRs as clinicians do not always document the care in the same way [38]. Such variance makes it difficult to extract the same information (e.g., outcome measures) consistently. Other data inconsistency issues such as missing data and selection bias also present significant challenges to researchers as data collection in real-world settings is usually heterogeneous and unstandardized. Second, most of the studies we identified focused on prediction or classification tasks and often overemphasized model performance rather than learning the casual effects [39,40]. Furthermore, most of these existing studies do not integrate *a priori* causal knowledge to guide the learning process, and as a result, no causal relationship can be estimated. Third, the transportability and interpretability of these studies also need to be further assessed. External validations using independent sources to ensure the findings are representative and generalizable are recommended, but such validation studies are often difficult to execute for multiple reasons: (1) sharing of individual-level clinical data remains difficult because of not only ethical and legal issues, but also market competition concerns; and (2) the lack of standardization and harmonization across the different data sources (e.g., inconsistent outcome measures), making replication studies unattainable.

Nevertheless, significant advancements have also been made to tackle these challenges. First, advances in AI methods, especially in deep learning, have prompted studies that consider heterogeneous data sources and types (e.g., clinical data, imaging, -omics data, and knowledge bases among others) in one coherent model. Li et al. developed a deep learning model based on recurrent neural networks to learn representation and temporal dynamics of longitudinal cognitive measures of individual subjects and combined them with baseline hippocampal Magnetic Resonance Imaging (MRI) measures to build a prognostic model of Alzheimer's disease dementia progression [41]. Other developments in deep learning include the ability to handle not only the temporal order of clinical events but also the long-term dependencies among the event as well as the time-varying effects of the covariates. For example, time-aware LSTM (T-LSTM) incorporate elapsed time information into the standard LSTM architecture to handle irregular time intervals in longitudinal EHR data [42] to learn disease subphenotypes. BEHRT, a new deep neural sequence transduction model for prediction of interpretable personalized risk using EHR data, models the temporal evolution of EHR data through utilizing various forms of sequential concepts and enabled the ability of incorporating multiple heterogeneous concepts (e.g., diagnosis, medication, measurements, and more) to further improve the accuracy of its predictions [43]. In NLP, new methods have been developed that can incorporate factual medical knowledge from existing ontologies/knowledge bases (e.g., the Unified Medical Language System) to further improve the performance of NLP tasks such as for clinical concept extraction[44].

Second, the use of causal modeling tools in AI, such as causal diagrams, could provide important additions to the implementations of causal inference using RWD. Causal modeling can also lead to improvements in the interpretability and adaptability of AI models in these drug development studies [45]. This concept of causal AI has been applied successfully in public health studies, such as the identification of occupational risk factors [46,47] and the prediction of diarrhea incidence in children [48] among others, and could be potentially used in future drug development research such as the "*target trial*"[49] framework aiming to establish causal treatment effects using RWD without conducting RCTs. Additionally, the emerging of explainable AI (XAI) could help to interpret and understand AI decisions. The XAI models use different mechanisms (e.g., feature interaction and importance, knowledge distillation, and rule extraction) on top of ML/DL models to generate interpretable outputs such as variable ranking [50], which ultimately help us understand why an AI system makes a certain decision. XAI models are particular useful for tasks such as drug repurposing because these tasks are generating hypotheses where plausible explanations are crucial.

Finally, the establishment of large research networks such as the national Patient-Centered Clinical Research Network (PCORnet) [51], Observational Health Data Sciences and Informatics (OHDSI) consortium [52], and the Clinical and Translational Service Award Accrual to Clinical Trails (CTSA ACT) network [53] facilitate the sharing of RWD. Each of these large networks consists of multiple sites across the US and internationally, and the same data infrastructure (i.e., the same ontologies and common data models) are being employed in each network. RWD from these networks represent a diverse set of patients and institutions and provide the opportunities to conduct large populational studies to understand factors that contribute to health and illness in a heterogeneous and real-world setting. In addition, deidentification strategies such as those for automated de-identification of massive clinical notes [54] have been widely applied to facilitate data sharing across different institutions. Furthermore, privacy-preserving record linkage tools have showed high precisions in linking and deduplicating patient records without sharing of protected identifiable information [55]. These deidentification strategies may not be applicable for every data type; nevertheless, they provide capabilities to facilitate data sharing across sites and integration of different data sources.

*Future applications*

There are several other scenarios where RWD and AI methods may be useful in the drug development process. For example, traditionally, clinical trial simulation (CTS) studies use computerized simulation methods on virtue populations to test different trial designs before resources are invested in conducting the actual clinical trial [56]. CTS that incorporates RWD can simulate its virtual populations more realistically. Furthermore, recent developments in the "*target trial*" framework—emulating hypothetical trials with RWD—enable us to identify unbiased initiation of exposures and reach an unbiased estimation of the casual relationships [49]. Combing the concept of modern trial emulation and the traditional CTS approaches, a trial simulation framework with RWD that can systematically test the different assumptions of a clinical trial to inform future trial design and produce causal results from RWD will be of high interest.

To facilitate the discovery of new drug targets, another emerging trend is the linkage of EHRs with other data sources such as biobanking data to study drug-phenotype and drug-gene interactions. For example, researchers from the Vanderbilt Electronic Systems for Pharmacogenomic Assessment (VESPA) Project [57], demonstrated that EHR-based biobanks could be cost-effective tools for establishing disease and drug associations as such applications allow the reuse of biological samples for multiple studies without incremental collection, extraction, or processing costs, and the integration with EHR system allows for centralized de-identification and phenotype annotations.

Finally, we would like to highlight the importance of the clinical and translational science lifecycle in the drug development process. For example, the drug repurposing signals identified from population-based studies will need to be looped back to the preclinical and clinical study stages for further validation and evaluation [58].

**Limitations of our work**
First, as a rapid review, our work is not comprehensive, but has provided a rapid and necessary summary and discussion of the topic. Second, our definition of AI is restricted to machine-/deep-learning methods (and their applications in NLP), and our definition of RWD is constrained to clinical data generated from the delivery of routine care (e.g., EHRs and claims data). Therefore, studies using AI methods such as automation and studies using data from personal devices such as social media and activity trackers were not included in our review. For example, social media data have shown promises in identifying AEs, although the noisy nature of social media data remains as a challenge [59,60]. These computational methods and data sources could provide additional insights into the drug development process and should be revisited in a future review.

**Concluding remarks**
The use of AI and real-world data has been emerging but focused on limited areas across several stages of the drug development process. Most AI studies focused on adverse events detection from clinical narratives in EHRs and a few studies explored applications for trial recruitment optimization and clinical drug repurposing. Benefitting from the detailed, longitudinal, multi-dimensional large collections of RWD and powerful AI algorithms, the use of AI methods on RWD provides golden opportunities in drug development, especially in identifying previously unknown associations and generating new hypotheses. Nevertheless, several current research gaps and challenges exist, such as issues in data quality, the difficult of sharing clinical data, and the lack of interpretability and transportability in AI models. We have highlighted examples of latest advancements in AI and data science that could address these challenges. For example, the increasing capability of deep learning models that can handle longitudinal and heterogeneous RWD and the raise of causal AI provide new research opportunities in drug development that can benefit from the combined use of AI and RWD.

**Acknowledgment**
None

21. Lai, Y. S., & Afseth, J. D. (2019). A review of the impact of utilising electronic medical records for clinical research recruitment. *Clinical Trials (London, England)*, *16*(2), 194–203. https://doi.org/10.1177/1740774519829709

22. Warrer, P., Hansen, E. H., Juhl-Jensen, L., & Aagaard, L. (2012). Using text-mining techniques in electronic patient records to identify ADRs from medicine use. *British Journal of Clinical Pharmacology*, *73*(5), 674–684. https://doi.org/10.1111/j.1365-2125.2011.04153.x

23. Luo, Y., Thompson, W. K., Herr, T. M., Zeng, Z., Berendsen, M. A., Jonnalagadda, S. R., Carson, M. B., & Starren, J. (2017). Natural Language Processing for EHR-Based Pharmacovigilance: A Structured Review. *Drug Safety*, *40*(11), 1075–1089. https://doi.org/10.1007/s40264-017-0558-6

24. Yang, X., Bian, J., Hogan, W. R., & Wu, Y. (n.d.). Clinical concept extraction using transformers. *Journal of the American Medical Informatics Association*. https://doi.org/10.1093/jamia/ocaa189

25. Fu, S., Chen, D., He, H., Liu, S., Moon, S., Peterson, K. J., Shen, F., Wang, L., Wang, Y., Wen, A., Zhao, Y., Sohn, S., & Liu, H. (2020). Clinical concept extraction: A methodology review. *Journal of Biomedical Informatics*, *109*, 103526. https://doi.org/10.1016/j.jbi.2020.103526

26. Yang, X., Bian, J., Gong, Y., Hogan, W. R., & Wu, Y. (2019). MADEx: A System for Detecting Medications, Adverse Drug Events, and Their Relations from Clinical Notes. *Drug Safety*, *42*(1), 123–133. https://doi.org/10.1007/s40264-018-0761-0

27. Zhang, H., Plutzky, J., Shubina, M., & Turchin, A. (2017). Continued Statin Prescriptions After Adverse Reactions and Patient Outcomes: A Cohort Study. *Annals of Internal Medicine*, *167*(4), 221–227. https://doi.org/10.7326/M16-0838

28. Spasic, I., Krzeminski, D., Corcoran, P., & Balinsky, A. (2019). Cohort Selection for Clinical Trials From Longitudinal Patient Records: Text Mining Approach. *JMIR Medical Informatics*, *7*(4), e15980. https://doi.org/10.2196/15980
12

**Figure 1**. The overall search and screening process.
**Figure 2.** Identified artificial intelligence (AI) and real-world data (RWD) applications across the different stages in the drug development process.
**Figure 3.** Number of original studies with artificial intelligence (AI) methods using real-world data (RWD) in the drug development process over the years.
**Figure 4.** Breakdown of real-world data sources, data types, and artificial intelligence (AI) methods used in the identified applications across the drug development process.  *Because of the overwhelming number of studies used AI-driven NLP methods, we separated NLP studies from other ML/DL studies.

**Tables**
**Table 1.** The main categories of artificial intelligence (AI) and real-world data (RWD) applications in drug development.

**Supplement**
**Supplement Table S1.** Search query and the complete list of keywords of literature search

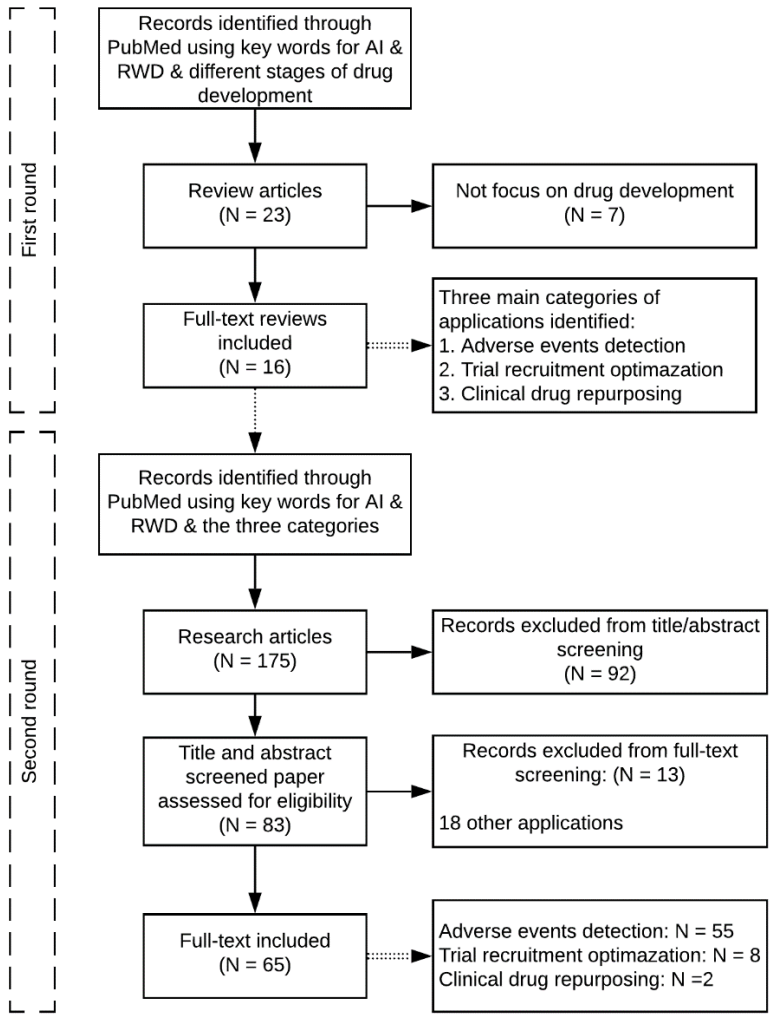

**Figure 1**. The overall search and screening process.

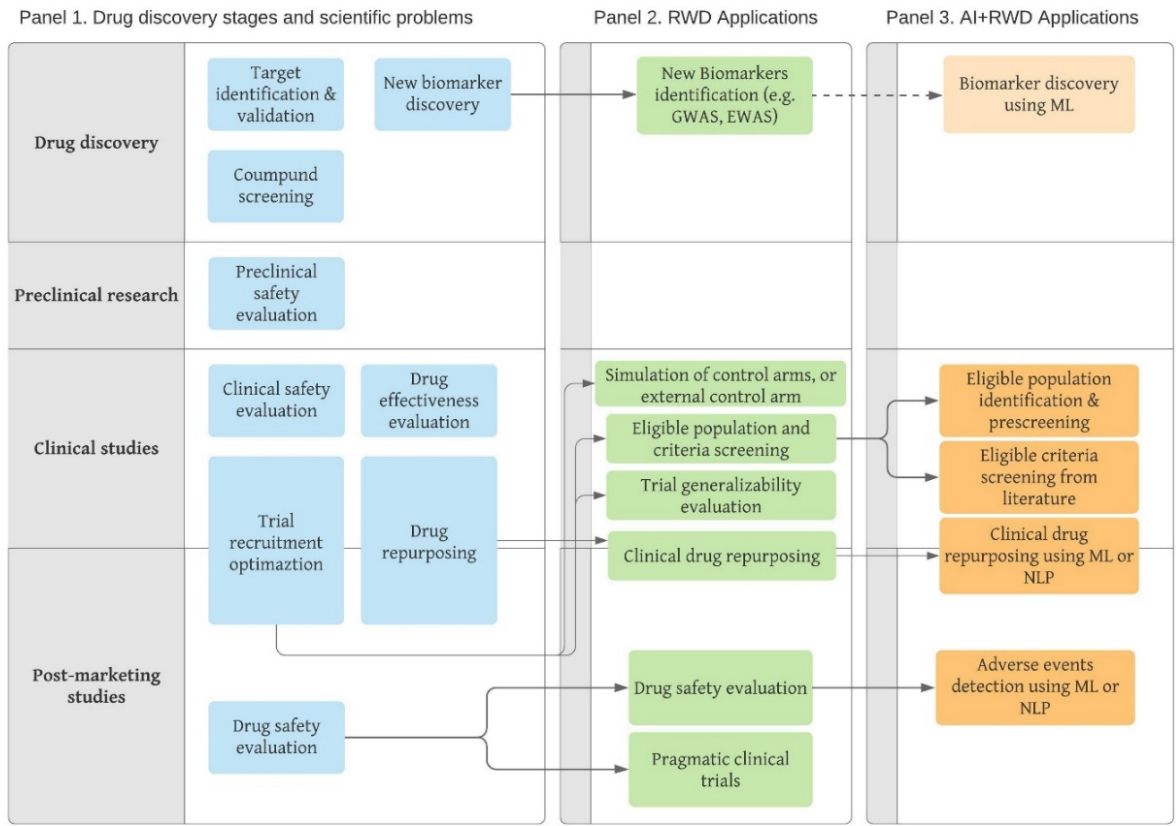

**Figure 2.** Identified artificial intelligence (AI) and real-world data (RWD) applications across the different stages in the drug development process.

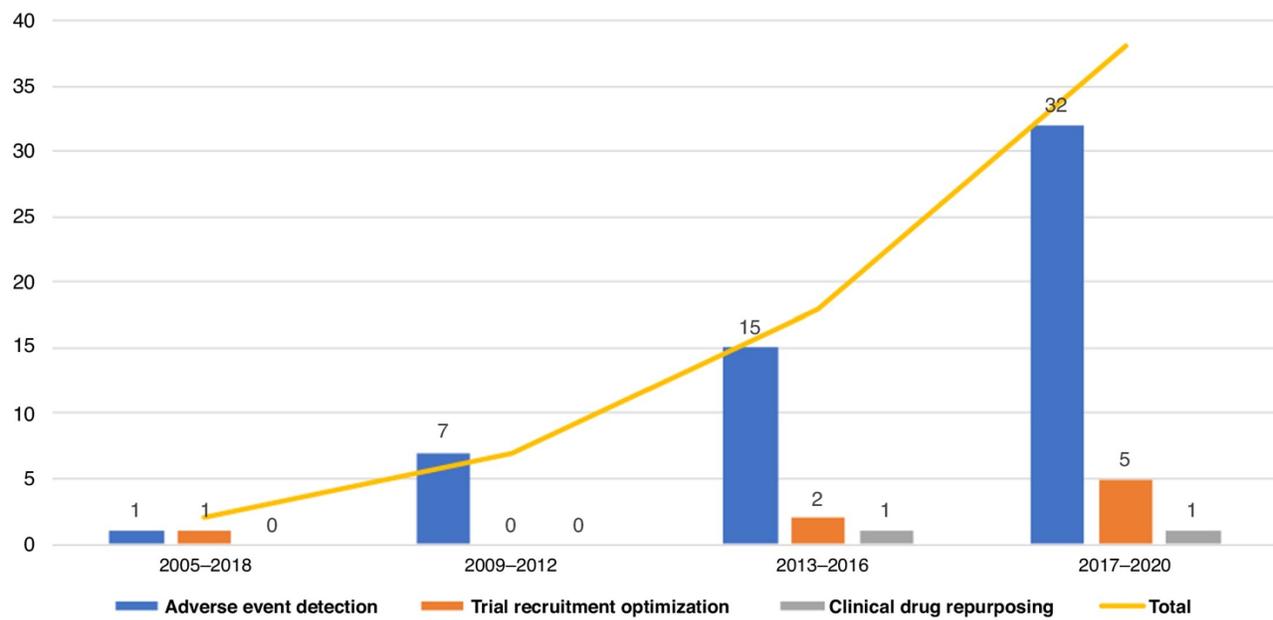

**Figure 3.** Number of original studies with artificial intelligence (AI) methods using real-world data (RWD) in the drug development process over the years.

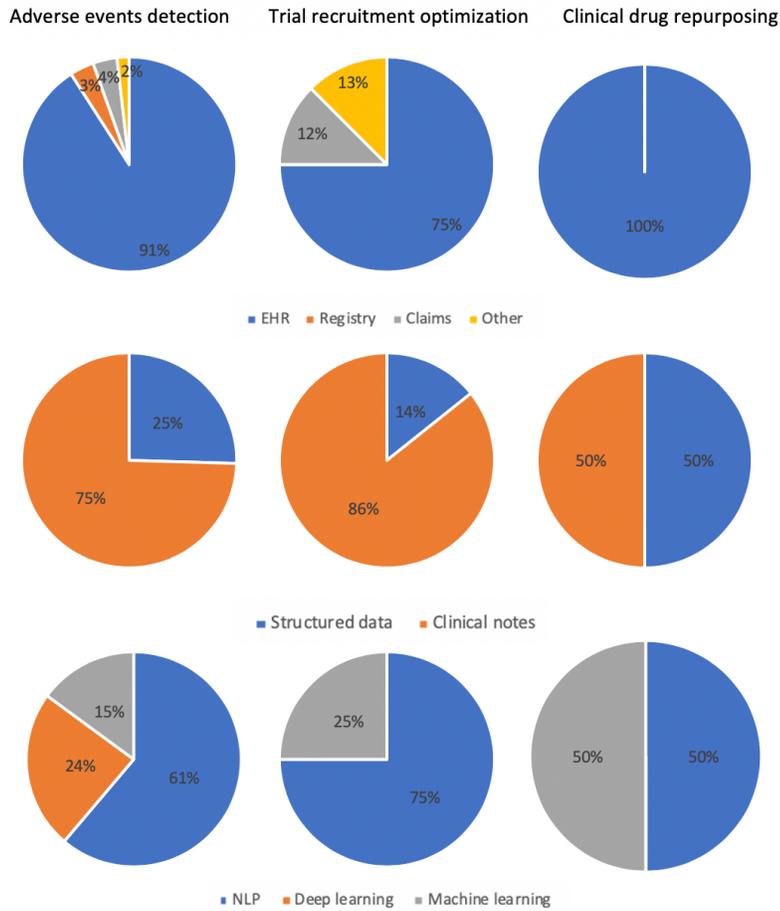

**Figure 4.** Breakdown of real-world data sources, data types, and artificial intelligence methods used in the identified applications across the drug development process.

**Table 1.** The main categories of AI and RWD applications in drug development.

| AI+RWD applications | Subcategories | Examples | Reference |
|---|---|---|---|
| Adverse events detection | Mining clinical notes using NLP | Deep learning-based NLP to detect adverse events in clinical notes extracted from EHR | [24] |
| | Mining structured EHR data | Predictive modeling of structured EHRs for adverse drug event detection | [25] |
| Recruitment optimization | Electronic recruitment through EHR | Electronic recruitment integrated into EHR workflow that sends electronic messages to recruit eligible patients | [26,27] |
| | Eligible population identification/pre-screening | Automated review of EHR to identify eligible population using NLP | [28] |
| Clinical drug repurposing | | Comparison between diabetic and non-diabetic cancer patients showed that the use of metformin was associated with decreased mortality after cancer diagnosis | [29] |
| NLP: natural language process; EHR: electronic health record | | | |